\input harvmac.tex      %% input your version
\input epsf.tex

\noblackbox

\def\omhats{ \hat{\omega}_{0,s} }
\def\omhatn{ \hat{\omega}_{0,n} }

%\input harvmac
%%%%%%%%%%%%%%%%%%%%%%%%%%%%%%%%%%%%%%%%%%%%%%%%%%%%%%%%%%%%%%%%%%%%%
%%%%
%%%%
%%%%
%%%%
%%%%
%%%%%%%%%%%%%%%%%%%%%%%%%%%%%%%%%%%%%%%%%%%%%%%%%%%%%%%%%%%%%%%%%%%%%
%%%%%%%
%%%%            Macros needed:  harvmac.tex 
%%%%
%%%%
%%%%%%%%%%%%%%%%%%%%%%%%%%%%%%%%%%%%%%%%%%%%%%%%%%%%%%%%%%%%%%%%%%%%%
%%%%%%%%%

%%%%%%%%%%%%%%%%%%% Figure commands  %%%%%%%%%%%%%%%%

%%%%%%%%%%%%%%%%%%%%%%%%%%%%%%%%%%%%%%%%%%%%%%%%%%%%%%%%%%%%%%%
%
%		DEFINITIONS FOR TEX
%
%%%%%%%%%%%%%%%%%%%%%%%%%%%%%%%%%%%%%%%%%%%%%%%%%%%%%%%%%%%%%%%
%
\def\bra#1{{\langle #1 |  }}

\def\bar{\overline}
\def\hat{\widehat}
\def\*{\star}
\def\[{\left[}
\def\]{\right]}
\def\({\left(}		
\def\){\right)}

%
%%%%%%%%%%%%%%%%%%%%%%%%%%%%%%%%%%%%%%%%%%%%%%%%%%%%%%%%%%%%%%%
%

\def\frac#1#2{{#1 \over #2}}
\def\inv#1{{1 \over #1}}

\def\d{\partial}

\def\ket#1{ | #1 \rangle}

\def\2pi{\hbox{$2\pi i$}}

\def\grad#1{\,\nabla\!_{{#1}}\,}
\def\dsl{\raise.15ex\hbox{/}\kern-.57em\partial}
\def\Dsl{\,\raise.15ex\hbox{/}\mkern-.13.5mu D}
%
%%%%%%%%%%%%%%%%%%%%GREEK LETTERS%%%%%%%%%%%%%%%%%%%%%%%%%%%%%%
%

\def\om{\omega}		
	
\def\vphi{\varphi}
%
%%%%%%%%%%%%%%%%%%%CALIGRAPHIC LETTERS%%%%%%%%%%%%%%%%%%%%%%%%%
%
\def\CA{{\cal A}}	\def\CB{{\cal B}}	
	\def\CE{{\cal E}}	
	\def\CH{{\cal H}}	
		
		\def\CO{{\cal O}}

\def\2pi{\hbox{$2\pi i$}}

\def\grad#1{\,\nabla\!_{{#1}}\,}
\def\dsl{\raise.15ex\hbox{/}\kern-.57em\partial}
\def\Dsl{\,\raise.15ex\hbox{/}\mkern-.13.5mu D}

\Title{ITP-97-072, hep-th/9706150}
{\vbox{\centerline{Eigenstates of the Atom-Field Interaction   }
\centerline{ and the Binding of Light in Photonic Crystals} }}

\bigskip
\bigskip

\centerline{Andr\'e Leclair\foot{On leave from Cornell University}}
\medskip\centerline{Institute for Theoretical Physics}
\centerline{University of California}
\centerline{Santa Barbara, CA 93106-4030}
\bigskip\bigskip

\vskip .3in

We solve for the exact atom-field eigenstates of a single atom in
a three dimensional spherical cavity, by mapping the problem onto
the anisotropic Kondo model.  The spectrum has a rich bound state
structure in comparison with models where the rotating wave
approximation is made.  It is shown how to obtain the Jaynes-Cummings
model states in the limit of weak coupling. 
  Non-perturbative Lamb shifts and decay
rates are computed.  The massive Kondo model is introduced to model
light localization in the form of photon-atom bound states in photonic
crystals.

\Date{6/97}
%\draftmode
%
%
%
%
%
%

%
%
%
%
%sample reference
%
% For the first time you ref an article:
%blah, blah, blah, see the paper\ref\ri{Author, .....} . 
%  The "\ri" is a label of the reference, so that when you
%   reference it later you just write  \ri.  
%
%sample equations
%  
%\eqn\one{
%\V 123 A\rangle_1 \vb_2 \vc_3 = 
%\langle h_1 \left[ V_A (0)\right] h_2 \left[ V_B (0) \right] 
%h_3 \left[ V_C (0) \right] \rangle . }
%
%  Above, \one is a label, so that later in the paper you just
% write  see equation \one.  
%
% Equations with multiple lines:
% \eqn\two{\eqalign{  a & = b \cr c & = d \cr}}
%
%
%  If you want equations  1a, 1b, 1c, separately labeled, etc:
%eq a,b,c etc
%\eqna\three
%$$\eqalignno{
%.............&\three {a} \cr
%..........&\three {b} \cr
%}$$
%\eqn\number{\eqalign{  ......... \cr}}       numbers automatically
%\newsec{Introduction }
%\centerline{Acknowledgments} 
%\figures
%\fig{1}{bla} 
%\appendix{A}{blaaaaaaaaaaaaa}
%\listrefs
%\end

\newsec{Introduction}

The system of two-level atoms coupled to radiation 
is a fundamental problem of physics, and much
work has been devoted to its study, especially in the context
of quantum optics\ref\rallen{L. Allen
and J. H. Eberly, {\it Optical Resonance and Two-Level Atoms}, 
Dover Publications, New York, 1987.}.     
For the most part, a satisfactory understanding of the problem
is obtained after making some  approximations,
most importantly the rotating-wave and slowly varying envelope
approximations. 
Though these approximations are well-justified near resonance,
it is important to have an understanding of what is missed in
making these approximations,  and to ascertain whether there are novel
effects of any significance.  

In this paper we present an exact solution to the problem of
a single two-level atom in a spherical cavity, without making
the above approximations.  This is a considerably more complicated
problem because of the vacuum fluctuations.   
The solution  is accomplished by mapping the optical problem onto the
Kondo problem of massless fermions interacting with a magnetic impurity,
as was done for a  one-dimensional 
fiber geometry in \ref\rLLLS{A. Leclair, 
F. Lesage, S. Lukyanov and H. Saleur,
{\it The Maxwell-Bloch Theory in Quantum Optics and the Kondo
Model},  hep-th/9701022.}.  

The novel features of our solution, which are unanticipated based
on previous studies, have to do with the spectrum of eigenstates of
the atom-field interaction.    For a single cavity mode in the
rotating-wave approximation, the atom-field eigenstates are known
from the solution of the Jaynes-Cummings model.  The parameters
of the latter model are the two-level splitting $\omega_0$,
and the coupling $\alpha$ of the atom to the radiation.  The latter
determines the decay rate $\Gamma_{\rm decay} = \alpha^2 L$ where
$L$ is the volume.  The eigenstates of the Jaynes-Cummings model
correspond to doublets of atomically dressed superpositions of
$N$ and $N-1$ photon states.   Our solution is given below  in terms of
a scattering theory description for a  spectrum of massless particles, 
and we describe how 
information about the atom-field eigenstates is encoded in 
the scattering matrices.     As a result, we find that 
the spectrum of  states
in our model is much richer, in a way that depends on the dimension-less
ratio $g =\Gamma_{\rm decay} / \pi \omega_0 $, which is a  quantum
coupling.    Roughly
speaking,  we find  that the  one-photon ($N=1$) eigenstates are 
$1/g$ in number, the lightest becoming identified with the Jaynes-Cummings
one at weak coupling (small $g$)  and near resonance, and the higher
ones arising as bound states, and as solitons.  
Interestingly, at the  
special strong coupling point 
$g=1/2$, the fundamental Jaynes-Cummings eigenstate becomes unbound
and disappears entirely from the spectrum, leaving only solitonic
states.  
In addition to these spectral 
features we are able to compute non-perturbative
expressions for the Lamb-shifted energy splittings and decay widths. 

The above  spectrum
has a sine-Gordon-like character.   Sine-Gordon theory in not unfamiliar
in this context, as it is well-known that a   resonant dielectric 
{\it medium}  of two-level atoms
in a semi-classical limit is well-described by the classical 
sine-Gordon theory.  This is the phenomena of 
self-induced transparency\ref\rsit{S. L. McCall
and E. L. Hahn, Phys. Rev. 183 (1969) 457.}. 
Our work can be interpreted as showing that the propagating spectrum
in the theory of self-induced transparency has an important significance
for a {\it single} atom in a cavity interacting with fully
quantized ratiation.  Our theory thus provides a bridge
between the theory of self-induced transparency and the Jaynes-Cummings
model.

In the final sections of the paper we apply our techniques to
the problem of an atomic impurity in a medium with a photonic
bandgap, with applications to photonic crystals in mind\ref\photonic{J. 
Joannopoulos, R. Meade, and J. Winn, {\it Photonic Crystals}, 
Princeton University Press, 1995; Nature, Vol. 396 (1997) 143.}.  
Photonic crystals, envisaged in early works of 
 Yablonovich and John\ref\ryab{E.
Yablonovich, Phys. Rev. Lett. 58 (1987) 2059.}\ref\rjohn{S. John, 
Phys. Rev. Lett. 58 (1987) 2486.}, are periodic dielectric structures
exhibiting gaps in the allowed energies of photon propagation, in
close analogy to electronic band structure.  Important effects are
expected when these materials are doped with atomic impurities.
Specifically, the spontaneous emission of an atom is severely
inhibited if the energy splitting of levels in the atom coincides
with an energy that is forbidden to propagate in the material.
Under these circumstances,  the light
can form a bound state with the atom, this photon-atom bound state
being the optical analog of an electron-impurity-level bound state in
the gap of a semiconductor\ref\rbind{S. John and J. Wang, Phys. Rev.
Lett. 64 (1990) 2418.}.  
In this paper we introduce a toy model where these effects can be
studied exactly by adding a mass term to the Kondo model in a way
the preserves the integrability of the latter.  Here, the origin
of the photonic bandgap is the gap between polariton branches. 
We solve this
massive Kondo model at a special point where it is equivalent to
a free fermion theory, and indeed find a photon-atom bound state
with the expected properties.  We can also compute exactly the
binding energy of this state.  
A different  integrable model of this phenomenon was 
studied in \ref\rrup{V. I. Rupasov and M. Singh, Phys. Rev. A 54 (1996) 
3614.}\ wherein the rotating wave approximation is made.

We present our results in the following way.  In section 2 we
use spherical symmetry to reduce the problem from three spatial
dimensions to one, and in section 3 describe the resulting
theory as a boundary quantum field theory.  For the resulting
theory to be exactly solvable, one needs to make an approximation
that favors photons in the vacinity of the resonance;  such
an approximation is not necessary in the one-dimensional case.
In section 4 we map
the problem onto the anisotropic Kondo model,  describe the
scattering spectrum in infinite volume, and present exact reflection
S-matrices expressed in terms of the 
Lamb-shifted resonant energies and decay widths.  
In section 5 we describe the meaning of topological
charge in the optical context.  In section 6 we describe how 
information about the atom-field eigenstates is encoded in the
reflection S-matrices by computing 
 the one-particle finite volume spectrum and comparing  this
to the Jaynes-Cummings spectrum. Here we show how for each particle
of the infinite volume spectrum one can associate polariton-like
states with optical phonon behavior at large and small energies. 
In section 7 the massive Kondo model is introduced to model
atomic impurities in photonic crystals.  The latter model is solved
at the special point $g=1/2$ in section 8, and this solution is
used in section 9 to study the localization of light.

\newsec{Spherical Dimensional Reduction} 

\def\omo{\omega_0} 
\def\rv{\vec{r}} 
\def\Ev{\vec{E}}
\def\Bv{\vec{B}}

We consider a two-level `atom', with spherically symmetric states
$|\pm \rangle$ which are eigenstates of the unperturbed atomic
hamiltonian $H_0^{\rm atom}$:
\eqn\eIi{
H_0^{\rm atom} ~ |\pm \rangle = \pm \frac{\omo}{2} |\pm \rangle, }
such that the energy splitting is $\omo$.  We work in spherical
coordinates $r,  \theta , \phi$, and place the atom at $r=0$.  
We will focus on electric-dipole transitions, thus we further
assume $\ket{-}$ is an angular momentum eigenstate 
$(l, m=0)$, whereas $\ket +$ is an  $(l+1, m=0)$ state.  
The wave-functions of these states are then
\eqn\eIii{
\bra{\rv} + \rangle = f_+ (r) Y_{l+1,0} (\theta, \phi ) ,
~~~~~
\bra{\rv} - \rangle = f_- (r) Y_{l0} (\theta, \phi ) .}

\def\om{\omega} 

In coupling the atom to radiation, one uses the electric-multipole
expansion as described in e.g. \ref\rjack{J. D. Jackson, 
{\it Classical Electrodynamics}, John Wiley and Sons, 1975.}.  
Expanding electric and magnetic fields in frequency
components, 
\eqn\eIiii{
\Ev (\rv , t) = \int d\omega ~ e^{-i\omega t} \Ev_\omega (\rv ) 
+ e^{i\omega t} \Ev_\omega^\dagger (\rv ) , }
and similarly for $\Bv$, and keeping only the electric-dipole term
in the multipole expansion, one has
\eqn\eIiv{\eqalign{
\Bv_\omega (\rv ) &= 2 \om^{3/2} ~ a(\om ) j_1 (\om r ) \vec{X}_{10} (\theta,
\phi ) \cr 
\Ev_\omega (\rv ) &= 2 i \om^{1/2} ~ a(\om ) 
{\vec{\grad{}}} \times j_1 (\om r ) \vec{X}_{10} (\theta,
\phi ) \cr }}
where 
$\vec{X}_{lm} =  \vec{L} Y_{lm} / \sqrt{l(l+1)}  $, and
$j_l$ is a spherical Bessel function. Using orthogonality
relations for $\vec{X}_{lm}$ and $j_l$, the free field 
hamiltonian is 
\eqn\eIv{
H_0^{\rm field} = \inv{8\pi} \int d^3 \rv \( \Ev^2 + \Bv^2 \) 
= \int_0^\infty d\om ~ \om a^\dagger (\om ) a (\om ) . }
In the quantum theory one has the commutation relation
\eqn\eIvi{
\[ a(\om ) , a^\dagger (\om') \] = \delta (\om - \om' ) . } 

\def\dv{\vec{d}} 

The atom is coupled to the radiation with the usual dipole interaction
$H_{int} = - \vec{d} \cdot \Ev $ where $\vec{d} $ is the electric
dipole operator of the atomic electron, $\dv = e \rv$.  
Letting $\rv = r \hat{r}$, one can show 
\eqn\eIvii{
\hat{r} \cdot \Ev_\om (\rv ) 
= - \sqrt{8 \om} ~ a(\om ) ~  \frac{j_1 (\om r) }{r} 
~ Y_{10} . }

We assume that the wavelength of relevant photons is large
compared to atomic dimensions, such that the $r$ dependence
in \eIvii\ can be pulled out of the matrix element:
\eqn\eIviii{
- \bra +  \dv \cdot \Ev_\om (\rv ) \ket - 
\approx \lim_{r\to 0} \( \frac{j_1 (\om r )}{r} \) 
\sqrt{8 \omega} ~ a (\om ) 
\bra + e r Y_{10} \ket - . } 

The matrix element $\bra + er Y_{10} \ket - $ can be related to the usual
reduced dipole moment $d$ defined as $\bra + d_m \ket - = d \delta_{m0} $,
where $d_m$ is the $m=\pm 1 , 0$ spherical component of the vector
$\dv$.  Using $\rv_m  =  r \sqrt{\frac{4 \pi}{3}} Y_{1m}  $, one
finds 
\eqn\eIx{
\bra +  e r Y_{10} \ket -  =  \sqrt{\frac{3}{4\pi}} d , }
where 
\eqn\edipole{ 
d =  e \frac{(l+1)}{\sqrt{(2l+3)(2l+1)}} 
\int_0^\infty dr r^3 f^*_+ (r) f_- (r) . }

Finally, using $\lim_{x\to 0} j_1 (x)/x = 1/3$, and the reality
of $Y_{l0}$, the complete hamiltonian is 
\eqn\eIxi{
H = H_0^{\rm field} + H_0^{\rm atom} + H_{int} 
}
where
\eqn\eIxii{
H_0^{\rm atom} = \frac{\omo}{2} S_3 , }
and 
\eqn\eIxiii{
H_{int} = \sqrt{\frac{2}{3\pi}} 
\Bigl\{ \int_0^\infty d\om \om^{3/2} \( a(\om ) e^{-i\om t} 
+ a^\dagger (\om) e^{i\om t} \) \Bigr\} 
\( d ~ S^+ + d^* ~ S^- \).}
Here, $S_3 , S^\pm$ are Pauli matrix operators:
\eqn\eIxiv{
[S_3 , S^\pm ] = \pm 2 S^\pm , ~~~~~[S^+ , S^- ] = S_3 . }

\newsec{Boundary Field Theory Description}

In this section we will give a boundary field theory
description of the model of the last section.  Boundary field
theories live on the half-line in space, $x\geq 0$, with non-trivial
interactions at the boundary.  
The half-line occurs automatically in the spherical reduction since
$r\geq 0$, and the boundary interaction is at $r=0$.     

The free hamiltonian $H_0^{\rm field}$ is equivalent to a free boson
on the half-line:
\eqn\eIIi{
H_0^{\rm field} =  \int_0^\infty dr  \inv{2} \( \( \d_t \phi \)^2 
+ \( \d_r \phi \)^2  \) , } 
with the supplemental Neumann boundary condition
$\d_r \phi (r=0,t) = 0$.  The mode expansion of $\phi$ can be 
written as 
\eqn\eIIii{
\phi (r,t) = i \int_{-\infty}^\infty \frac{dk}{\sqrt{2\pi}} 
\inv{\sqrt{2|k|}} 
\( a(k) e^{-i\vec{k} \cdot \vec{x} } 
- a^\dagger (k) e^{i \vec{k} \cdot \vec{x} } \) , }
where 
$\vec{k} \cdot \vec{x} = |k|t - kr$. 
The Neumann boundary condition enforces $a(k) = a(-k )$, so that
at $r=0$, $\phi (r,t)$ can be expressed as an integral over 
$k>0$ modes.  

In order to make the
problem solvable, as we will see,  we make an  
 approximation that
favors  photons in the vacinity of $\omega_0$, 
so that $\omega^{3/2}$ in \eIxiii\ is
replaced by $\omega_0 \sqrt{\omega}$. 
Then, 
\eqn\eIIiii{
\int_0^\infty d\omega  ~ 
\omega^{3/2} \( a(\omega) e^{-i\omega t} + a^\dagger (\omega) e^{i\omega t}
\)  \approx  \sqrt{\pi} \omega_0 \d_t \phi (0,t) .  }
If $d$ is complex we absorb the phases of $d, d^*$ into the definition
of $S^\pm$ without changing the commutation relations \eIxiv.  Finally,
we obtain the hamiltonian\foot{This hamiltonian is closely related to
the so-called Lee model and its generalizations\ref\rlee{T. D. Lee,
Phys. Rev. 95 (1954) 1329; O. W. Greenberg and S. S. Schweber,
Nuovo Cimento 8 (1958) 378;  I. Bialynicki-Birula, Nucl. Phys. 12
(1959) 309.}, except that in our model  the
fermion fields are fixed at one spacial location.  The latter two 
papers in \rlee\ study the renormalization 
problems associated with what amounts to
the rotating-wave approximation.}  
\eqn\eIIiv{
H = H_0^{\rm field} + \frac{\omo}{2} S_3 
+ \frac{\beta}{4} \d_t \phi (0,t) \( S^+ + S^- \), }
where 
\eqn\eIIv{
\beta = \sqrt{ \frac{32}{3} } d \omo . } 
To lowest order in perturbation theory, the hamiltonian \eIIiv\ gives
the well-known result  $\Gamma_{\rm decay}  = \frac{4}{3} d^2 \omega_0^3$.   
Our solution below presents all-order $\beta^2$ 
corrections to this result based on
\eIIiv.   

Though  we need to make the approximation
\eIIiii\ in order to solve the model,  
fortunately 
in a one-dimensional fiber geometry, one finds $\omega^{1/2}$
rather than $\omega^{3/2}$ in \eIxiii, so that the model is 
exactly solvable without any further approximations\rLLLS.  
All of the subsequent results of this paper hold in  
a fiber geometry with 
\eqn\efiberb{
\beta_{\rm fiber} = \( \frac{16\pi}{\CA_{eff}} \)^{1/2} d , }
where $\CA_{eff}$ is the effective cross-sectional area of the fiber. 

The parameter $\beta^2 /8\pi $ is a dimension-less coupling
constant, where the strong coupling regime is large $\beta^2$. 
Using the lowest order perturbative result for
$\Gamma_{\rm decay}$, this coupling can also be expressed as 
\eqn\ecoupling{
g \equiv 
\frac{\beta^2}{8\pi}  \approx \inv{\pi}  \frac{\Gamma_{\rm decay}}{\omo},
 ~~~~~~~(g \ll 1). }
In order to get some idea of the size of the coupling for real
atoms, we can consider
a hydrogen-like atom with the nuclear charge $Ze$.  The largest value 
of $\beta$ occurs for the $1s$ to $2p$ transition, where
$g  = Z^2 e^6 2^{11} 3^{-9} / \pi \approx 
Z^2 ~ 10^{-8} $. 
Here, $g$ goes as $e^6$ since the Coulomb interaction determines both
$\omega_0$ and $d$.  One can perhaps hope for larger values of $g$ 
in artificial atoms, e.g. quantum dots, 
 wherein  $d$ and $\omega_0$ are fixed by distinct  
physics.

\newsec{Mapping to the Kondo Model} 

\subsec{The Exact Scattering  Spectrum}

In \rLLLS, 
a single atom in a fiber, which is 
described by a hamiltonian of the form \eIIiv\ except that the
theory lives on the full line $-\infty < x < \infty $ and
$\beta$ is different, was mapped onto the Kondo model.  
Here, the result is even simpler since one doesn't have to 
fold the system.  
Namely, $H$ is related by a unitary transformation $U$ to the
bosonized form of the anisotropic Kondo hamiltonian $H_K$: 
\eqn\eIIIi{
H= U^\dagger H_K U , }
where 
\eqn\eIIIii{
H_K = H_0^{\rm field}  
+ \frac{\omo}{2} \( S^+ e^{i\beta \phi (0) /2 } 
+ S^- e^{-i\beta \phi (0) /2 } \) , }
and 
\eqn\eIIIiii{
U = \inv{\sqrt{2} } e^{i \beta S_3 \phi (0) /4 } 
\( S_3 + S_+ + S_- \) . } 

Since $H$ and $H_K$ are related by a unitary transformation,
the quantum mechanics can be formulated in either the 
`optical picture' or the `Kondo picture'.  Matrix elements
of operators $\CO$ in the two pictures are simply related:
\eqn\eunit{
\langle \psi' | \CO | \psi \rangle_{\rm optical} 
= {}_K \langle \psi' | \CO_K | \psi \rangle_K , }
where 
$|\psi \rangle_K = U |\psi \rangle$, and 
$\CO_K = U \CO U^\dagger$.  
We remark that the trivial atomic hamiltonian $H_0^{\rm atom}$ 
in the optical picture becomes the interaction in the Kondo
picture.  

\def\Gam{\Gamma_{\rm decay}}
\def\omhat{\hat{\omega}_0}

Both models \eIIiv\eIIIii\ have ultra-violet divergences, 
and furthermore the quantum operator $U$ is in need of 
regularization.   We will take the point of view that 
the equation \eIIIi\ defines the regularization of $H$ once
we have regularized $H_K$ and $U$.  The regularization of 
$H_K$ is as in the sine-Gordon model and amounts to properly
normal-ordering the exponential operators;  this leads to the  anomalous
scaling dimension $-g$ for the parameter $\omega_0$.  
It can be checked that this is consistent with the renormalization
performed directly in the optical hamiltonian $H$ as was done in 
\ref\rfibrill{A. LeClair, Phys. Rev. A56 (1997) 782.}, where it
was shown that to lowest order the beta function reads 
$\mu \d_\mu \omega_0 = -g \omega_0$.

The physical parameters of the optical problem are $g$, which
is governed by the strength of the dipole coupling, and the
energy scale set by the two-level splitting.  The parameter
$\omo$ in \eIIIii\ is a bare, unphysical parameter; physical
energy scales are a function of $\omo$, $g$ and an ultraviolet
cut-off $\mu$.  In the scattering theory solution presented below, 
physical energy scales will be set by the parameter $\omega_B$.
This scale is a known function which we will not need, but 
for completeness include from \rLLLS:
\eqn\ewB{
\omega_B = \inv{\sqrt{\pi}} \cot \( \frac{\pi g}{2-2g} \) 
\frac{ \Gamma \( \frac{1-2g}{2-2g} \) }
{\Gamma \( \frac{2-3g}{2-2g} \) }
\( \frac{\omo (\mu) }{2} \Gamma (1-g) \)^{\inv{1-g}}
 . }

In the Kondo model the physical parameters are the dimension-less
anisotropy parameter $g$ and the `Kondo-temperature' $T_K$, which
is related to $\omega_B$ as follows:
\eqn\eTk{
T_K = \tan \( \frac{\pi g}{2-2g} \) \omega_B . } 
$T_K$ is defined such that the atomic impurity contribution to the
partition function behaves as 
\eqn\ekondoT{
Z_{\rm atom} = 2 \cosh \( T_K /T \) + \CO (g) , } 
thus
$T_K$ is convenient for describing thermodynamical properties,
as was done in \rLLLS.
For  small $g$,  $T_K \approx \omo /2$;  
this is consistent with the fact that when $g=0$, $Z_{\rm atom}$ should
represent the partition function for a two-level system with energies
$\pm \omo/2$. 
Henceforth, we will express all energies in terms of $\omega_B$.

Most importantly, the spectrum of $H$ and $H_K$ are identical.  
Thus we can infer what are the  quantum states that exactly diagonalize
the atom-field interaction from knowledge of the exact spectrum
in the Kondo model\foot{We refer the reader to 
\ref\rfend{P. Fendley, Phys. Rev. Lett. 71 (1993) 2485.}\ref\rsal{F.
Lesage, H. Saleur and S. Skorik, Nucl. Phys. B474 (1996) 602.} for
a description of the Kondo model that is the most useful in  our
context.  An exact Bethe-ansatz solution of the Kondo model was
found in \ref\rbethe{N. Andrei, K. Furuya and J. Lowenstein, 
Rev. Mod. Phys. 55 (1983) 331\semi A. M. Tsvelick and P. B. Wiegmann,
Adv. Phys. 32 (1983) 453.}.  References to other earlier papers
can be found in \rsal. }.  
     This spectrum consists of a rich spectrum
of massless  particles which have a sine-Gordon-like 
character.  The origin of the sine-Gordon spectrum is due to the
fact that a bulk sine-Gordon interaction is compatible, as far
as integrability goes, with the boundary interaction. (See below.)  
This is analogous to the treatment of the boundary sine-Gordon theory
in \ref\rfendl{P. Fendley, H. Saleur and N. Warner, Nucl. Phys. B 430
(1994) 577.}, the difference being in the reflection S-matrices.   
Namely, the spectrum consists of a soliton and anti-soliton, and
$[1/g - 2]$ breathers which can be viewed as soliton-anti-soliton
bound states.  The meaning of this spectrum will be elaborated upon
in section 5.    The point $g=1/2$, the so-called Toulouse point of
Kondo physics, 
 can be formulated as a free fermion theory using bosonization. 
(See below.) 
In the repulsive regime $g \geq  1/2$, there are only
solitons and anti-solitons.   
For  $g > 1$, the interaction is
an irrelevant operator and the theory breaks down.  
In the bulk these particles propagate with a massless dispersion
relation:  
\eqn\edisp{
\eqalign{
E_a = P_a = \mu_a  ~ e^\theta   ~~~~~~~&{\rm for ~ right-movers} \cr
E_a = - P_a = \mu_a  ~ e^{-\theta} ~~~~~~~&{\rm for ~ left-movers} \cr 
, }}
where $a  \in \{ 1,2, ..< (1-g)/g, s, \bar{s} \}$ 
 is an index running over the breathers and solitons, and $\theta$ is
a rapidity variable. 
The parameters $\mu_a$ are related as follows:
\eqn\ebreath{
\mu_n = 2 ~ \mu~ \sin \( \frac{n \pi g}{2-2g} \) , ~~~~~~
\mu_s = \mu_{\bar{s}}  = \mu,  }
where $\mu$ is an arbitrary, unphysical energy scale which can be set to
$1$.

The interactions of these particles with the atom are encoded in 
the exact reflection S-matrices for these particles at the
boundary\rfend\rsal. 
Let $R_{a}^{b} (\theta ) $ denote the S-matrix for particle of type $a$
to scatter off the boundary and be reflected into a particle of type
$b$.  
Defining $\theta_B$ as:
\eqn\ethB{
\omega_B = \mu e^{\theta_B}, }
the 
explicit expressions for the reflection S-matrices for the
left-movers  are 
\eqn\ereflect{\eqalign{
R_+^- (\theta )  &=  R_-^+ (\theta) = \tanh \( \frac{\theta + \theta_B}{2} 
- \frac{i\pi}{4} \) \cr
R_+^+ (\theta) &= R_-^- (\theta)  = 0 \cr
R_n^n (\theta) &= 
\frac{\tanh \( \frac{\theta + \theta_B }{2} - \frac{i\pi g n}{4(1-g)} \) } 
{\tanh \( \frac{\theta + \theta_B }{2} + \frac{i\pi g n}{4(1-g)} \) } 
.\cr }}
Here $+ = s$ and $- = \bar{s}$. 

\def\ohatn{\hat{\omega}_{0,n}}

It will be useful to express the reflection S-matrices in terms of
resonances $\ohatn$,   widths $\Gamma_n$,  and physical energies $E$:
\eqn\eIIIxiib{\eqalign{
R_+^- (E) &= \frac{ \omhats^2 + E^2} {\omhats^2 - E^2 + i \Gamma_s E }
\cr
R_n^n (E) &= \frac{ \omhatn^2 - E^2 - i \Gamma_n E}
{\omhatn^2 - E^2 + i \Gamma_n E }, 
\cr }}
where
\eqn\eIIIxiid{\eqalign{
\omhats &= \omega_B , ~~~~~\Gamma_s = 2 \omega_B \cr
\omhatn &= 2 \sin \( \frac{n\pi g}{2-2g} \) \omega_B , ~~~~~
\Gamma_n = 4 \sin^2 \( \frac{n\pi g}{2-2g} \) \omega_B . 
\cr }}
Near the resonances $E \approx \omhatn$, the reflection S-matrices
have a Lorentzian signature: 
\eqn\eIIIxiie{
R_n^n \approx - \frac{ (\Gamma_n/2)^2}{ (\omhatn - E)^2 + (\Gamma_n /2)^2 }
.}
For the solitons, one has 
\eqn\eIIxiif{
R_+^- \approx -i \frac{ (\Gamma_s/2)^2}{(\omhats - E)^2 + (\Gamma_s /2)^2 }
.}
As we will show in  section 5,  the quantities 
$\omhatn , \Gamma_n$ determine the energy spectrum of eigenstates in finite
volume. 

It is well known in the quantum sine-Gordon theory
literature  that the $n=1$ breather is the particle corresponding
to the scalar field $\phi$ itself\ref\rZZ{A. B. Zamolodchikov
and Al. B. Zamolodchikov, Ann. Phys. 120 (1979) 253.}.   
To make this more explicit, 
in \rLLLS\ electric field correlation functions were computed in the
single 1-breather approximation which is remarkably good for values
of $g$ up to $\sim 1/5$.  Because of \eIIIxiie, the vacuum power
spectrum is Lorentzian where $\hat{\omega}_{0,1} $ and $\Gamma_1$ 
represent the Lamb-shifted two-level atomic energy splitting
and decay rate.  These agree with lowest order perturbative computations:
\eqn\eapprox{\eqalign{
\hat{\omega}_{0,1} &= \omega_0 + \CO(g) \cr 
\Gamma_1 &= g\pi \omega_0 + \CO(g^2) . \cr }}

\newsec{Topological Charge} 

In the usual bulk sine-Gordon theory on the full line, the
soliton and anti-soliton carry topological charge $Q=\pm 1$,
where 
\eqn\eQ{
Q = \frac{\beta}{2\pi} \int_{-\infty}^\infty dx ~ \d_x \phi 
= \frac{\beta}{2\pi} \( \phi (\infty) - \phi (-\infty) \) . }
In our boundary version of the problem, solitons are still characterized
by a charge $Q = \beta ( \phi (r=\infty ) - \phi (r=0 ) )/2\pi$. 
The value of $\phi$ 
at $r=0$ isn't fixed, thus topological charge conservation
can be violated.  
For simplicity, set $\phi (r=\infty ) = 0$.  An incoming soliton
$(Q=1)$ with $\phi (r=0) = -2\pi/\beta$ can be reflected at $r=0$
into an outgoing anti-soliton with $\phi (r=0) = 2\pi /\beta$, 
violating charge conservation by 2 units.  In this process, the
field at the boundary interpolates between $-2\pi/\beta$ and 
$2\pi/ \beta$.  
The fact that $R_\pm^\pm = 0$ means that only processes which
violate topological charge conservation by two units  
are allowed in the soliton sector. 

We can relate the topological charge of solitons to simple properties
of the electric field $E\propto \d_t \phi$.  Imagining the soliton
as a pulse localized in space, then an incoming soliton at time
$-t$  and an outgoing
anti-soliton at time $t$, where $t=0$ corresponds to complete
absorption,   have pulse profiles in space simply related by 
$\phi \to -\phi$.  Thus, the phase of the electric field of the outgoing
anti-soliton is shifted by $\pi$ in comparison to the incoming
anti-soliton. 
 
Consider a soliton scattering process where the atom is initially in its
ground state.  When the soliton is absorbed, the atom is precisely
in its excited state;  it then emits an anti-soliton and returns to its ground
state in the far future.  
This can be seen by examining
the original optical states $|\pm \rangle$ in the Kondo picture. 
Defining $|\pm , \phi (0) \rangle_K = U |\pm \rangle$, one has
\eqn\etop{
|\pm , \phi (0) \rangle_K = \inv{\sqrt{2}} 
\( e^{i\beta \phi (0)/4 } |+\rangle 
~ \pm ~ e^{-i\beta \phi (0) /4 } |- \rangle \) . }
Note that 
\eqn\eevolve{
|- , \phi (0) + \frac{2\pi}{\beta} \rangle_K = i |+, \phi (0) \rangle_K ,
~~~~~
|- , \phi (0) + \frac{4\pi}{\beta} \rangle_K = - |-, \phi (0) \rangle_K .
} 
Thus, a soliton incident on the atom in its ground state in the far
past corresponds to $|-, -2\pi/\beta\rangle$, and when the soliton
is completely absorbed $\phi (0) = 0$ and the state has evolved to 
$i|+, 0 \rangle$, i.e. is in its excited state.  In the far future,
$\phi (0) = 2\pi/ \beta$, and the state is $-|-, 2\pi /\beta \rangle$. 
The atom has necessarily returned to its ground state due to energy
conservation.  Note however that the atomic state has changed phase
by $\pi$.  

The fact that soliton absorption excites that atom to precisely its
excited state indicates that the solitons are the most fundamental
excitations.  As we will see below, at strong coupling only the 
solitons remain in the spectrum. 

\newsec{Optical Phonon Spectrum and the Jaynes-Cummings Model Limit}

Though the scattering theory described above provides
 an exact solution of the model, its precise meaning is not entirely
clear from what we have done so far.  The original Hilbert space,
$\CH$, consisting of the two-level space tensored with the 
photon Hilbert space, has been replaced with an infrared 
description $\CH_{\rm particle}$. 
The latter Hilbert space contains a spectrum
of particles with continuous energies, and does not have a two-level
structure.   
The distinguishing feature of this spectrum of particles is that they provide
a basis that diagonalizes the atom-field interaction;  the existence of
the  two-level
atomic Hilbert space is encoded in the reflection S-matrices $R$. 
A similar treatment of the simpler case of harmonic-oscillator 
atomic impurities was given in \ref\rkon{R. Konik and A. LeClair, 
{\it The Scattering Theory of Oscillator Defects in an Optical Fiber}, 
 hep-th/9701016.}.  
Irrespective  of these remarks, 
the free field hamiltonian $H_0^{\rm field}$ is a linear theory in vacuum,
so an intrinsic understanding of the non-linear sine-Gordon-like spectrum
seems called for. 

It is also desirable  to describe eigenstates of the atom-field
interaction in  a fashion that is more understandable in terms of
the original Hilbert space $\CH$.   This is a difficult problem
in general, since it requires computing the inner-product of states
in $\CH$ with states in $\CH_{\rm particle}$.  
Exact eigenstates  can be described in the space $\CH$ for the
much simpler Jaynes-Cummings (JC) model.  In this section we describe
how information about the atom-field  eigenstates is encoded 
in   the scattering
theory description, and indeed we show how to  obtain the known
 JC eigen-energies in the
appropriate limit.   This exercise  leads to the understanding that
for each particle of the spectrum of section 3 there is an associated
optical phonon, in a sense that will be explained.

\def\omc{\omega_c}

\subsec{Jaynes-Cummings Model}

The Jaynes-Cummings model\ref\rjc{E. T. Jaynes and 
F. W. Cummings, Proc. IEEE {\bf 51} (1963) 89. } 
 is an exactly solvable model of the atom-field
interaction which arises under various approximations to the hamiltonian
\eIIiv.   Let us suppose that the one-dimensional volume of the
cavity is $L$.  The first approximation is to only consider a single  
discrete photon mode of the cavity of frequency $\omc$ near
$\omega_0$.    Namely,
the free photon Hilbert space is built from the single mode 
creation-annihilation operators  
$a \equiv a(\omc )$ and $a^\dagger = a^\dagger (\omc )$ satisfying
$[a , a^\dagger ] = 1$.  
Next, we make the rotating-wave approximation and drop the terms
$a S^- $ and $a^\dagger S^+$ in \eIIiv.  Finally, we drop the
exponentials $\exp(\omc - \omega_0)$ in the interaction.
 Recalling that in 
finite volume $\int dk / \sqrt{2\pi} \to \inv{\sqrt{L}} \sum_k $, 
one obtains the hamiltonian:
\eqn\ejci{
H = \omc \> a^\dagger a + \frac{\omo}{2} S_3 + \alpha 
\( a S^+ + a^\dagger S^- \) , }
where
\eqn\ejcii{
\alpha = \sqrt{ \frac{\pi g \omc}{L}  } . }
The volume $L$ drops out of many physical quantities; for
instance $\Gamma_{\rm decay} = \alpha^2 L|_{\omc = \omo}
 = \beta^2 \omo /8 $. 
In the limiting case of a micro-cavity of size tuned to the
resonant frequency $ \omc = \omega_0$ with $\omo/\pi = 1/L$, one has
$\alpha_{\rm micro} = \sqrt{g\pi} \omo $.  

The above hamiltonian is easily diagonalized exactly.  Let
$|N, \pm>$ denote a basis of states with $N$ photons 
and $\pm$ the two-level states of the atom:
\eqn\ejciv{
a^\dagger a |N , \pm \rangle  = N |N, \pm \rangle 
, ~~~~~
S_3 |N , \pm \rangle = \pm | N, \pm \rangle . } 
The hamiltonian has a $2\times 2$ block diagonal form  where
the blocks act on the pairs of states $|N-1, + \rangle, 
|N , - \rangle $.  Each block is easily diagonalized leading
to an energy spectrum $\CE^\pm_N$, $N=1,2,...$:
\eqn\ejcv{
\CE_N^\pm = \omc (N-1/2) \pm \inv{2} 
\sqrt{ (\omc - \omo)^2 + 4 \alpha^2 N } , } 
where the eigenstates $H \Psi^\pm_N = \CE_N^\pm \Psi^\pm_N$ are given by
\eqn\ejcvi{\eqalign{
\Psi_N^+ &= \cos \Theta |N-1, + \rangle + \sin \Theta |N , - \rangle \cr
\Psi_N^- &= -\sin \Theta |N-1, + \rangle + \cos \Theta |N , - \rangle , \cr
}}
where
\eqn\ejcvi{
\tan \Theta = \frac{
\sqrt{ (\omc-\omo)^2 + 4\alpha^2 N } + (\omc - \omo ) }
{\sqrt{ (\omc-\omo)^2 + 4\alpha^2 N } - (\omc - \omo ) }
.}
This spectrum is observable 
experimentally\ref\rnorm{R. J. Thompson, G. Rempe and
H. J. Kimble, Phys. Rev. Lett. 68 (1992) 1132.}.

\subsec{Optical Phonon Spectrum}

Let us describe how to obtain the JC spectrum from our scattering
theory under certain limits.  In order to compare with the 
single cavity mode approximation, we must place the atom inside
a finite volume, letting the spherical cavity have a diameter $L$. 
There is now a quantization condition on allowed momenta which 
depends on the reflection S-matrices.  Consider a wave-function for
a particle from the spectrum of section 3  as it traverses the 
radius of the cavity, is reflected and returns to the edge.  
This wave-function picks up a phase which must equal unity:
\eqn\efvi{
e^{i P L } R(E) = 1. }
Taking the logarithm, and using $P= -E$, one obtains
\eqn\efvii{
E + \frac{i}{L} \log R (E) = \frac{2\pi m}{L} , }
where $m$ is some integer.  The quantity on the RHS determines the
mode of the cavity: $\omega_c = 2\pi m/L$.  

Let us now make two approximations.   As in the JC model we assume
the cavity is tuned near resonance $\omega_c \approx \omhatn$ for the
quantization of the n-th breather with $R= R_n^n$ in \efvii.  
Secondly, at weak coupling, for the low $n$ breathers, 
it is safe to suppose $\Gamma_n \ll |\omhatn -E |$, since
$\Gamma_n \approx n^2 \pi^2 g^2 \omega_B$ and  $\omhatn 
\approx n\pi g \omega_B $.  Under these approximations, 
$\log R_n^n \approx -i \Gamma_n /(\omhatn - E) $.  Inserting this
into \efvii, one obtains a quadratic equation for $E$, the solution
for the n-th breather being: 
\eqn\efviii{
E^\pm_{N=1, n} (\omega_c ) = 
\frac{\omhatn + \omega_c}{2}   \pm \inv{2}  
\( (\omega_c - \omhatn )^2 + 4 \Gamma_n / L \)^{1/2} . }
Multiparticle ($N>1$) eigenstates are characterized by more 
complicated quantization conditions involving the bulk S-matrix\rfendl.  
Since the ground state energy of the JC model is $-\omo/2$, whereas
that of the scattering description is set to zero, one should subtract
$\omo/2$ from the above energies in order to compare.  Using the
small $g$ limits, 
\eqn\efviv{
E^\pm_{N=1, n} (\omega_c ) - \omega_0 /2 
\approx  \frac{(n-1)}{2}  \omo + \omega_c \pm \inv{2}  
\( (\omega_c - n \omega_0 )^2 + 4 n^2 g\pi \omega_0 / L \)^{1/2}  . }
From \ejcii, with $\omega_c \approx \omega_0$, one sees that for the 
$n=1$ breather, one-particle energies correspond precisely to the 
one-photon JC energies: 
\eqn\efviv{
E^\pm_{N=1, n=1} - \omo/2 = \CE^\pm_{N=1} . }

At $\omega_c = \omega_0$, the $N=1$ JC eigenstates are symmetrically
split about $\omega_0 /2$.  Let us define then a scaled energy
splitting $E'$ such that:
\eqn\esplita{
\frac{\CE^\pm_1 - \omega_0 /2}{\omega_0} = \pm E' = \pm (g/2)^{1/2} . }
We can then compare this with the `exact' prediction 
\efvii.   Setting $E' = (E-\hat{\omega}_{0,1})/\hat{\omega}_{0,1}$ 
in \efvii\ where
$R$ is the reflection S-matrix for the first breather $R_1^1$, 
and $\omega_c = \hat{\omega}_{0,1} = 
2\pi/L$, one finds that $E'$ is a solution of 
the equation 
\eqn\esplit{
E' + \frac{i}{2\pi} 
\log \( 
\frac{E'^2 + 2 E' + 2 i \sin(\pi g/(2-2g)) (E' + 1) }
{E'^2 + 2 E' -  2 i \sin(\pi g/(2-2g)) (E' + 1) }
\) = 0. }
The differences between the JC model and the result
\esplit\ are displayed in figure 1.  At small $g$ they of course
agree very well.  The strong departure from the JC model at 
$g=1/2$ is due to the fact that this is the threshold for the
disappearance of this particle.  

%%%%%%%%%%%%%%%%%%%  fig 1  %%%%%%%%%%%%%%%%%%%%%
\midinsert
\epsfxsize = 3in
\bigskip\bigskip\bigskip\bigskip
\vbox{\vskip -.1in\hbox{\centerline{\epsffile{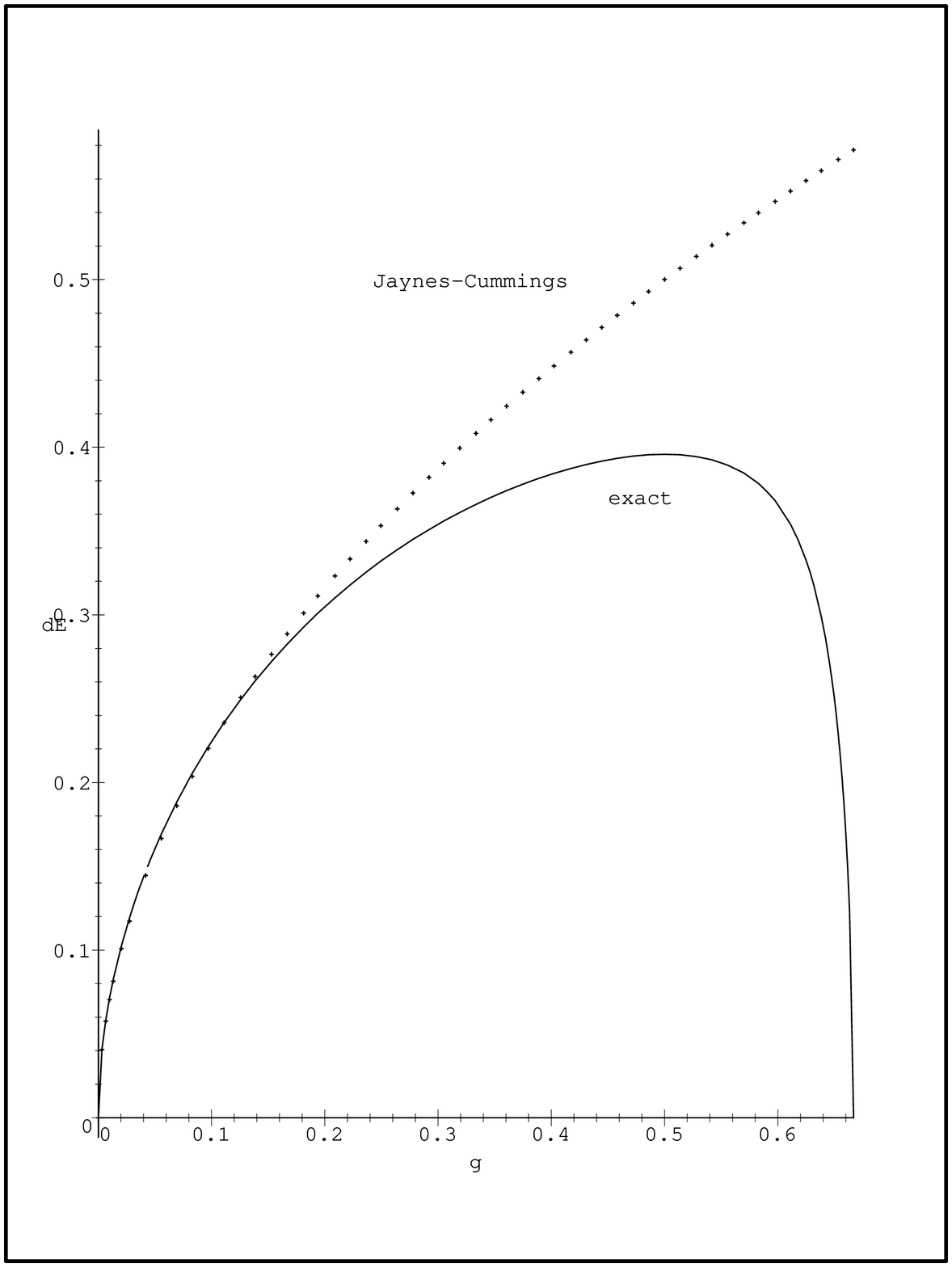}}}
\vskip .1in
{\leftskip .5in \rightskip .5in \noindent \ninerm \baselineskip=10pt
Figure 1.
Comparison of the energy splitting of the $N=1$ photon Jaynes-Cummings
eigenstates and the predictions of our model.  The y axis is 
$E'$, which,   for the JC model
 is defined in \esplita, whereas for our model it is a solution
to \esplit.   
\smallskip}} \bigskip
\endinsert
%%%%%%%%%%%%%%%%%%% End fig 1   %%%%%%%%%%%%%%%%%%%%%%%

The higher $n$-breather $N=1$ particle states are not predicted
by the JC model.   We believe 
this is because higher resonances $E\approx n \omo$ 
are eliminated in the rotating wave approximation.  
Based on the above observations we can formulate the following appealing
picture of the spectrum.  At large volumes, the  energies behave as
\eqn\epol{\eqalign{
&E^+_{N=1, n} \approx \omega_c , ~~~~~~ E^-_{N=1, n} \approx \omhatn , 
~~~~~ \omega_c \gg \omhatn , \cr
&E^+_{N=1, n} \approx \omhatn , ~~~~~~ E^-_{N=1, n} \approx \omega_c , 
~~~~~ \omega_c \ll \omhatn . \cr
}}
Thus the $E^+$ branch is optical phonon-like at small $\omega_c$ 
and photon-like at large $\omega_c$, whereas for $E^-$ this is reversed. 
For $n=1$, these states are the precursor to the well-known polariton
branches which occur in the context of a dielectric medium of atoms.
(See for instance \ref\mermin{N. W. Ashcroft and N. D. Mermin,
{\it Solid State Physics}, Chapter 27, W. B. Saunders Co., 1976.} and 
the next section.)
Here, there is no medium, however 
these `vacuum-polaritons' are characterized by the regimes where
they are phonon-like, so we will simply refer to them as optical
phonons.  The $n=1$ breather optical phonon is the basic excitation
of the JC model. 
Since the n-th breather leads to an optical phonon at energy 
approximately $n\omega_0$ for small $g$,  the latter can be understood
as a bound state of $n$ optical phonons of energy $\omo$.  

We thus propose that our model contains a rich spectrum of 
optical phonons of energy $\omhatn$ , $\omhats$, in correspondence
with the breather/soliton spectrum in section 3.  The n-th optical 
phonon is a bound state of $n$ fundamental $n=1$ phonons.  As $g$ 
increases the higher optical phonons become unbound and disappear
from the spectrum one by one.  When $g= 1/(n+1)$, from \eIIIxiid\ one
sees that $\omhatn = 2 \omhats$.  Thus as $g$ increases and reaches
the point $1/(n+1)$, the n-th optical phonon unbinds into two
solitonic phonons.  This
also implies that all of the n-th breather optical phonons can
also be viewed as a bound state of two solitonic optical phonons. 
Finally, when $g=1/2$, the $n=1$ optical phonon, which is the fundamental
excitation of the JC model,  also disappears from the spectrum leaving
only the solitonic phonons for all $g\geq 1/2$.

\newsec{Atomic Impurity in a Photonic Crystal }

\def\vep{\varepsilon}

In this section we place the atom in a dielectric medium.  One
can generally model such a medium with a dielectric constant
\eqn\eIVi{
\varepsilon (\omega) = \varepsilon_\infty 
+ (\varepsilon_\infty - \varepsilon_0 ) 
\frac{ \omega_T^2 }{\omega^2 - \omega_T^2 } , } 
giving the dispersion relation $\varepsilon (\omega ) = k^2 / \omega^2 $. 
(See for instance \mermin.)  
For simplicity, let us set $\vep_\infty = 1$.  
The dispersion relation has two polariton branches 
$\omega_\pm (k)$ as shown in figure 2.  Since 
$\omega_+ (k=0) = \sqrt{\vep_0} \omega_T$, and 
$\omega_- (k=\infty) = \omega_T$, there is a gap between the
two branches $E_{gap} = (\sqrt{\vep_0} - 1 ) \omega_T$. 
Though this is not the manner in which photon bandgaps are thought to
arise in e.g. \photonic, we believe it serves as a good model for
the physics we are trying to study.  

%%%%%%%%%%%%%%%%%%%  fig 2  %%%%%%%%%%%%%%%%%%%%%
\midinsert
\epsfxsize = 3in
\bigskip\bigskip\bigskip\bigskip
\vbox{\vskip -.1in\hbox{\centerline{\epsffile{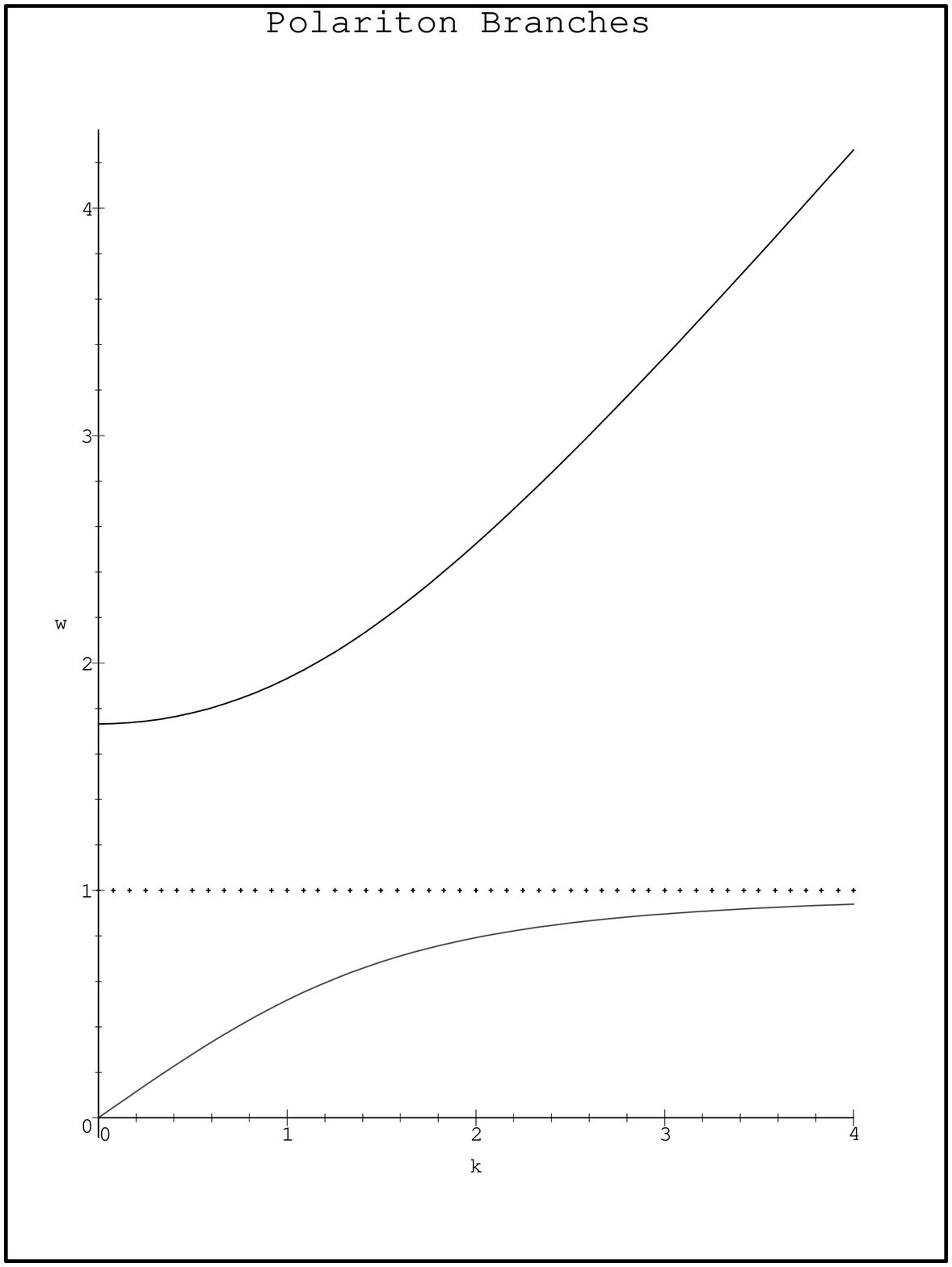}}}
\vskip .1in
{\leftskip .5in \rightskip .5in \noindent \ninerm \baselineskip=10pt
Figure 2.
The two polariton branches of the dispersion relation following from
\eIVi, with $\vep_0 = 3$.  
Both axes are in units of $\omega_T $.  The upper branch
intersects the $y$-axis at $\sqrt{\varepsilon_0}$. 
\smallskip}} \bigskip
\endinsert
%%%%%%%%%%%%%%%%%%% End fig 2   %%%%%%%%%%%%%%%%%%%%%%%

\def\mgap{m_{\rm gap}}

In a certain limit, the dispersion relation is approximately
relativistic.  Namely, let us treat the lower branch as 
constant $\omega_- (k) = \omega_T$, and shift the energy of the
upper branch $\omega_+ = \omega_T + \omega'$.  Then, if 
$\vep_0$ is large, $w'$ is large compared to $\omega_T$ and
$\omega'^2 - k^2 \approx \mgap^2$ with  
\eqn\emass{
\mgap = \sqrt{\vep_0 -1} ~ \omega_T   .} 
For $\vep_0 \gg 1$, $ \mgap \approx E_{gap}$. 
Alternatively, one can view this limit as taking $\omega_T \to 0$,
and $\vep_0 \to \infty$, keeping  $\mgap = \sqrt{\vep_0} ~ \omega_T$
fixed.  In this limit the lower branch disappears and the upper
branch is a massive relativistic dispersion relation.

One can model this gap in the medium while preserving the integrability
by adding a sine-Gordon interaction in the bulk:
\eqn\eIVii{
H_0^{\rm field} \to H_0^{\rm field} - \frac{m^2}{\beta^2} 
\int _0^\infty dr  \cos (\beta \phi ) . } 
Choosing $m$ as in \emass, the above interaction gives the photon
the proper dispersion relation.  
Mapping onto the Kondo model as in \eIIIi, one now obtains
\eqn\eIViii{
H^{(m)}_K =    
\[ \int_0^\infty dr  \( \inv{2} (\d_t \phi )^2 
+ \inv{2} (\d_r \phi )^2 - \frac{\mgap^2}{\beta^2} \cos (\beta \phi ) \) \]   
+ \frac{\omo}{2} \( S^+ e^{i\beta \phi (0) /2 } 
+ S^- e^{-i\beta \phi (0) /2 } \) . }
The desirable feature of the theory \eIViii\ is that the bulk and
boundary interactions are compatible as far as integrability is
concerned.  Namely, the exact bulk sine-Gordon spectrum also diagonalizes
the boundary interaction with the atom.  The spectrum of the model
is the same as in section 3, except that the particles are now
massive with a mass scale determined by $\mgap$.  
The dispersion relation is now
\eqn\massive{
E_a = M_a \cosh \theta , ~~~~~~ P_a = M_a \sinh \theta , }
satisfying $E_a^2 - P_a^2 = M_a^2$. 
The masses $M_a$  are related as in \ebreath, 
\eqn\emassmass{
M_n = 2 M_s \sin \frac{n \pi g}{2-2g},}
but now the soliton 
mass $M_s$ is a physical parameter depending on $\mgap$. 
Since the first breather is identified with the photon, we will set
\eqn\efix{
M_1 = \mgap  .} 
 
In the literature on the 
Kondo model, a bulk mass term as in \eIViii\ is never considered 
since one deals with massless fermions near
the Fermi surface.   
We believe the model \eIVii\ is integrable since the bulk massive
spectrum is compatible with the spectrum that diagonalizes the
boundary interaction.  The boundary reflection S-matrices for
the model \eIViii\ are unknown.  
For simplicity, in this paper we will limit ourselves to the
model at $g=1/2$ and derive the S-matrices in the next section.  
We intend  to present the general case elsewhere. 

\newsec{Free Fermion Point:  Massive Case}

Remarkably, at the point $g=1/2$ in the strong coupling regime,
the theory is equivalent to a free fermion theory. 
Though this is well-known in the Kondo physics literature,
this is a novel phenomenon in the optical context since 
at this value of the coupling the perturbatively dressed photon, 
in the sense of the Jaynes-Cummings model, 
disappears entirely 
from the spectrum.  The theory at this coupling 
  will be considered in greater detail
in \ref\rbassi{Z. Bassi, R. Konik, and A. Leclair, in preparation.}. 
Here, we consider the bulk massive case, which has not been
studied before,   with applications to 
photonic crystals in mind.     

\subsec{Physical Parameters} 

We first define the physical parameters.  When $g=1/2$, 
$T_K$ diverges
 however $\omega_B$ remains finite\foot{
One finds $\omega_B \propto \omo^2$, which is consistent 
with the anomalous $\sqrt{\rm mass}$ dimension of $\omo$.}. 
For $g=1/2$ we define the physical parameters   as 
\eqn\eVi{
\omhat = 2 \hat{\omega}_{0,s} = 2 \omega_B  , ~~~~~~~
\Gam = 2 \omhat . } 
Note that this precisely 
corresponds to the 1-breather expressions 
 $\hat{\omega}_{0,1}$
and $\Gamma_1$ evaluated at  
$g=1/2$.  The parameter $\omhat$ represents the energy splitting
of the atom; note that since this is twice $\omhats$, the
soliton is in resonance with {\it half} of the energy splitting.  

From \emassmass\ one sees that the breather mass $M_1$ is
exactly twice the soliton, or fermion, mass $M_s$, which
reflects the fact that the lowest breather is a soliton/anti-soliton
bound state and just becomes unbound at $g=1/2$.  
In this section, we will express everything in terms of the 
fermion mass $M \equiv M_s = \mgap/2$.

\subsec{Reflection S-matrices}

We now derive the reflection S-matrices in a treatment that 
 is similar to what was done for the boundary
sine-Gordon theory in \ref\rAKL{M. Ameduri, R. Konik, and 
A. Leclair,  Phys. Lett. B354 (1995) 376.}. 

\def\vphi{\varphi}
\def\vphib{\bar{\vphi}} 
\def\psib{\bar{\psi}}

In the massless limit, the scalar field decomposes into left
and right-moving components
$\phi = \vphi (z^+ ) + \vphib (z^-)$, where
$z^\pm = t\pm r$. 
The Neumann boundary condition $\d_r \phi = 0$ then reads
$\d_{z^+} \vphi (z^+ ) = \d_{z^-} \vphib (z^-) $ at
$r=0$.   This implies
\eqn\eVbi{
\vphi (z^+) = \vphib (z^-) - \frac{\sigma}{\sqrt{4\pi}}, }
where $\sigma$ is some constant.  

The fermion fields, with topological charge $\pm 1$, are the following:
\eqn\eVbii{
\psi_\pm = e^{\pm i \sqrt{4\pi} ~\vphi} , 
~~~~~
\psib_\pm = e^{\mp i \sqrt{4\pi} ~\vphib} . }
In terms of the fermions, the Neumann boundary condition reads 
\eqn\eVbiii{
\psi_\pm = e^{\mp i \sigma} ~ \psib_\mp, ~~~~~~~~~(r=0)}
which breaks the topological charge symmetry.  
The action which enforces the Neumann boundary condition is 
\eqn\eVbiv{\eqalign{
S_{\rm free} = \int_0^\infty dr dt 
& \[ i \psi_- (\d_t - \d_r ) \psi_+ + i \psib_- (\d_t - \d_r ) \psib_+ 
- M (\psib_- \psi_+ + \psi_- \psib_+ ) \] 
\cr 
&  
-i \int dt \( e^{i\sigma} \psi_+ \psib_+ + e^{-i\sigma} \psi_- \psib_- \) .
\cr
}}

Using \eVbi, the interaction can be written as
\eqn\eVbv{
S_{\rm int} = - \frac{\lambda}{2} \int dt 
\[ (\psi_+ a_- + \psib_- a_+ ) S_+ 
+ S_- (a_+ \psi_- + a_- \psib_+ ) \] ,}
where $a_\pm = e^{\mp i \sigma/2}$.  
The parameter $\lambda $, with units of 
$\sqrt{\rm mass}$,  will be related to $\omega_B$ below. 
The $a_\pm$ should be regarded  as fermionic operators:
\eqn\eVbvb{
a_\pm \psi = - \psi a_\pm,} 
for $\psi \in ( \psi_\pm, \psib_\pm )$.  
The complete action is $S = S_{\rm free} + S_{\rm int}$. 

The boundary equations of motion which follow from varying the 
action with respect to the fermion fields are, at $r=0$:
\eqn\eVvi{\eqalign{
\psi_+ - e^{-i\sigma} \psib_- &= \frac{i\lambda}{2} S_- a_+, 
~~~~~~~~~ \cr
\psib_+ - e^{-i\sigma} \psi_- &= \frac{i\lambda}{2}  a_+ S_+,  
~~~~~~~~~ \cr
}}
and the hermitian conjugate of these.   
The time derivative of the operators on the RHS of \eVvi\ are 
determined by their commutation relations with the hamiltonian,
$\d_t \CO = -i [ \CO, H]$.  Using \eVbvb\ and 
$\{ S_+ , S_- \} = 1$, one obtains
\eqn\eVvii{\eqalign{
\d_t ( S_- a_+ ) &= \frac{i\lambda}{2} \( \psi_+ + a_+^2 ~ \psib_- \) \cr
\d_t (  a_+  S_+ ) &= -\frac{i\lambda}{2} \( \psib_+ + a_+^2 ~ \psi_- \) \cr
}}
and their hermitian conjugates (at $r=0$).  Combining \eVvi, 
\eVvii\ and their hermitian conjugates, one obtains two 
independent equations at $r=0$:
\eqn\eVviii{\eqalign{
\d_t \( \psi_+ - e^{-i\sigma} \psib_- \) &= 
- \frac{\lambda^2}{4} \( \psi_+ + e^{-i\sigma} \psib_- \) \cr
\d_t \( \psib_+ - e^{-i\sigma} \psi_- \) &= 
 \frac{\lambda^2}{4} \( \psib_+ + e^{-i\sigma} \psi_- \) .\cr
}}

The reflection S-matrices can be easily computed from \eVviii. 
The fermion fields have the mode expansions
\eqn\eVix{\eqalign{
\psi_+ (r,t) = \sqrt{\frac{M}{4\pi} } 
\int_{-\infty}^\infty d\theta ~ e^{-\theta/2} 
\( A_- (\theta ) e^{-i k\cdot r} - A_+^\dagger (\theta) e^{i k\cdot r} \)
\cr
\psib_+ (r,t) = \sqrt{\frac{M}{4\pi} } 
\int_{-\infty}^\infty d\theta ~ e^{+\theta/2} 
\( A_- (\theta ) e^{-i k\cdot r} + A_+^\dagger (\theta) e^{i k\cdot r} \),
\cr
}}
with $\psi_- = \psi_+^\dagger , ~ \psib_- = \psib_+^\dagger$.  
The variable $\theta$ is the rapidity:
$k\cdot r = M \cosh (\theta) t - M \sinh (\theta) r $,
and the $A, A^\dagger$ are fermion annihilation/creation operators
satisfying $\{ A_\pm (\theta) , A^\dagger_\pm (\theta') \} 
= \delta(\theta - \theta')$.  

The scattering of the particles off the boundary at $r=0$ can be
formulated  by formally introducing a boundary 
operator $\CB$
satisfying the algebraic 
relation\ref\rghosh{S. Ghoshal and A. Zamolodchikov, Int. J. Mod. Phys.
{\bf A9} (1994) 3841.}
\eqn\eVx{
A_a^\dagger (\theta) ~\CB = R_a^b (\theta) A_b^\dagger (-\theta) ~ \CB.}
Separating the integral over $\theta$ in \eVix\ into $\theta <0$
and $\theta >0$, and making a change of variables $\theta \to -\theta$ in
in the former, one obtains
\eqn\eVxi{\eqalign{
&\[ - e^{-\frac{\theta}{2}} \( i M \cosh \theta + \frac{\lambda^2}{4} \) 
A_+^\dagger (\theta) 
- e^{-i\sigma} e^{\frac{\theta}{2} } 
\( i M \cosh \theta - \frac{\lambda^2}{4} \) A_-^\dagger (\theta) \] \CB 
\cr 
&~~~~~~~= 
\[  e^{\frac{\theta}{2}} \( i M \cosh \theta + \frac{\lambda^2}{4} \) 
A_+^\dagger (-\theta) 
+ e^{-i\sigma} e^{-\frac{\theta}{2} } 
\( i M \cosh \theta - \frac{\lambda^2}{4} \) A_-^\dagger (-\theta) \] \CB 
\cr
&~
\cr
&\[  e^{\frac{\theta}{2}} \( i M \cosh \theta - \frac{\lambda^2}{4} \) 
A_+^\dagger (\theta) 
- e^{-i\sigma} e^{-\frac{\theta}{2} } 
\( i M \cosh \theta + \frac{\lambda^2}{4} \) A_-^\dagger (\theta) \] \CB 
\cr 
&~~~~~~~= 
\[  -e^{-\frac{\theta}{2}} \( i M \cosh \theta - \frac{\lambda^2}{4} \) 
A_+^\dagger (-\theta) 
+ e^{-i\sigma} e^{\frac{\theta}{2} } 
\( i M \cosh \theta + \frac{\lambda^2}{4} \) A_-^\dagger (-\theta) \] \CB 
.\cr
}}
From this one can read off the reflection S-matrices: 
\eqn\eVxii{\eqalign{
R_+^- (\theta ) &= e^{- i \sigma} 
\frac{\sinh \theta}{ \cosh (\theta - \gamma(\theta))} ,
~~~~~~~~
R_-^+ (\theta ) =  - e^{ i \sigma} 
\frac{\sinh \theta}{ \cosh (\theta - \gamma(\theta))} 
\cr
R_+^+ (\theta ) &=  R_-^- (\theta) =  -  
\frac{\cosh \gamma(\theta) }{ \cosh (\theta - \gamma(\theta))}, \cr
}}
where
\eqn\eVxiii{
e^{\gamma(\theta)} = 
\frac{  \cosh \theta  - i \Delta}
{  \cosh \theta +  i \Delta  } ,
~~~~~~~~~~~ \Delta = \frac{\lambda^2}{2\mgap} 
.}
The parameter $\sigma$ is unphysical since it simply corresponds to
a phase, hence we set it to zero.  

One can easily check that $R$ satisfies the unitarity and 
crossing symmetry constraints\rghosh:
\eqn\eVxiv{\eqalign{
R_a^b (\theta) R_b^c (-\theta) &= \delta_a^c \cr
R_{\bar{a}}^b \( \frac{i\pi}{2} - \theta \) 
&= 
- R_{\bar{b}}^a \( \frac{i\pi}{2} + \theta \), \cr
}}
where $\bar{a} = -a$.  

One can also confirm that these reflection matrices have the proper
massless limit.  For the left-movers, the massless limit is obtained
by letting $\theta \to \theta - \alpha$, and taking 
$\alpha \to \infty$, $M\to 0$ while keeping 
$M e^\alpha /2 = \mu$ held fixed.  In this way, the massless 
dispersion relation \edisp\ is recovered from the massive one
\massive.  Taking this limit in $R$ one finds
\eqn\eVxv{\eqalign{
R_\pm^\mp  &\to  \pm 
\tanh \(  \frac{\theta + \theta_B}{2} - \frac{i\pi}{4} \), \cr
R_\pm^\pm &\to 0, \cr
}}
with
\eqn\eVxvi{
\omega_B = \mu e^{\theta_B} = \frac{\lambda^2}{4} . }
This agrees with \ereflect\ up to unphysical phases.   
The relation \eVxvi\ can now be used to express everything in
terms of physical parameters.  In particular, \eVxvi\ combined with
\eVi\ gives
\eqn\edelta{
\Delta = \frac{\omhat}{\mgap} . } 

In terms of the physical energy $E$, 
\eqn\eRE{\eqalign{
R_+^- (E) &= \frac{ \sqrt{E^2 - \mgap^2/4 }}{E} 
\( 
\frac{ E^2 + \omhat^2/4}{E^2 - \omhat^2 /4 + i \omhat \sqrt{E^2 - \mgap^2/4}}
\) 
\cr
R_+^+ &= - \frac{\mgap}{2E} 
\( 
\frac{ E^2 - \omhat^2/4}{E^2 - \omhat^2 /4 + i \omhat \sqrt{E^2 - \mgap^2/4}}
\) . \cr
}}

\newsec{The Binding of Light to Atoms} 

Using results from the last section, we now
 study the binding of light to atoms in 
photon bandgap materials in the simplest case of the free
fermion point. 
Recall the physical parameters are the renormalized 
two-level splitting $\omhat$, and the bandgap of the medium
$\mgap$.  

Light localized at, or bound to, the atom corresponds to a 
`boundary bound state' of the kind described in \rghosh.  
Boundary bound states exist if there are poles in the
reflection S-matrices $R_a^b (\theta)$ in \eVxii\ on the
physical strip $0\leq Im~ \theta \leq \pi$.   Poles in $R$
occur when $\cosh (\theta - \gamma(\theta)) = 0$.  Letting
$\theta = i u$,  poles occur at solutions of the equation
\eqn\eVIi{
\sin^2 u + 2 \Delta \sin u + \Delta^2 - 1 = 0.} 
Requiring the pole to be on the physical strip leads to a single
boundary bound state at $u= u_b$ satisfying
\eqn\eVIii{
\sin u_b = 1-\Delta \geq 0. }
Thus there exists a light-atom bound state if $\Delta \leq 1$, or
\eqn\eVIiib{
\omhat \leq \mgap.} 

The necessary condition \eVIiib\ for the existence of a bound state
is simple to understand:  the atom normally decays spontaneously
by emitting photons of energy $\omhat$, however such a photon
cannot propagate in the medium and thus becomes bound to the atom.  

The  energy of the boundary bound state above the 
ground state energy  is $M \cos u_b $, and this
represents the binding energy:    
\eqn\ebind{
E_{\rm bind} = \( \frac{\mgap \omhat}{2}  ~ 
\( 1- \frac{\omhat}{2\mgap} \) \)^{1/2} . } 
The bound state is stable if the binding energy is less than the 
mass of the lightest particle, $E_{\rm bind} < M$, or 
$\omhat < \mgap/2$.  

The existence of this bound state can be used to design a filter for
light near the binding energy.  Recall that with no atomic impurities,
light with energy $E< \mgap$  is forbidden to propagate in the medium.
The bound state is at an energy $E_b < \mgap/\sqrt{2}$ due to the 
inequality \eVIiib, thus it is within the gap.  
If there are many atomic impurities, 
one can imagine that   light of energy near $E_b$ can resonate with
the bound state and in principle  still propagate by tunneling along
the impurities. 
One indication of this is the enhanced reflection amplitude for 
energies near $E_b$: 
\eqn\ehop{
R_+^- (E) \approx \frac{i}{E-E_{\rm bind}}  
~ \frac{ (\mgap - \omhat)^2 \omhat}{4 E_b^2}
. }

One can determine the reflection S-matrices for particle scattering
off of the bound state itself\rghosh.  In this simple case 
they turn out to be identical to the reflection S-matrices $R_a^b$ 
for the scattering off the ground state computed in section 4.

\newsec{Conclusions} 

We have found an exact solution to the problem of eigenstates of
the atom-field interaction in a spherical cavity which shows a rich
spectrum in comparison with models based on the rotating wave approximation.
The spectrum has solitons and bound states, in spite of the fact that
there is no medium, i.e. 
we are dealing with a single atom in vacuum.  
In a sense our work shows that the well-known solitons of classical
non-linear optics have a counterpart for a single atom in vacuum.  

By introducing a bulk mass term into the Kondo model in a way that
preserves integrability we have shown how one can study analytically
the occurance and properties of photon-atom bound states in photonic
crystals.  

It is interesting to explore the possibility of utilizing the full spectrum
of eigenstates in novel device applications.  The solitonic mode
being in a sense the most fundamental, one can imagine 
a `soliton laser' based on the existence of this mode.

\bigskip\bigskip

\centerline{\bf Acknowledgments}

I wish to thank Z. Bassi, R. Konik, A. Imamoglu, F. Lesage, V. Rupasov, 
 H. Saleur, M. Stone  and X.-G. Wen  for discussions. 
This work is supported by the National Science foundation, in part
through the National Young Investigator Program, and under
Grant No. PHY94-07194.

\listrefs

\bye